\documentclass[11pt]{article}
\usepackage{epsfig}
\usepackage{amsmath}

\newcommand{\nn}{\nonumber}
\newcommand{\et}[1]{e^{\mbox{\small $#1$}}}
\newcommand{\lra}{\leftrightarrow}
\newcommand{\dt}{\! \cdot \!}
\newcommand{\wdg}{\! \wedge \!}
\newcommand{\crs}{\! \times \!}
\newcommand{\half}{{\textstyle \frac{1}{2}}}
\newcommand{\sig}{\sigma}
\newcommand{\gam}{\gamma}
\newcommand{\del}{\delta}
\newcommand{\Del}{\Delta}
\newcommand{\eps}{\epsilon}
\newcommand{\gi}{\gamma_{1}}
\newcommand{\gj}{\gamma_{2}}
\newcommand{\gk}{\gamma_{3}}
\newcommand{\go}{\gamma_{0}}

\newcommand{\deriv}[2]{\frac{\partial #1}{\partial #2}}
\newcommand{\sigr}{\sig_r}
\newcommand{\sigth}{\sig_\theta}
\newcommand{\sigph}{\sig_\phi}
\newcommand{\dr}{\partial_r}

\newcommand{\om}{\omega}
\newcommand{\alp}{\alpha}
\newcommand{\lam}{\lambda}
\newcommand{\Rrev}{\tilde{R}}

\newcommand{\dphi}{\partial_\phi}

\newcommand{\dift}{\partial_t}

\newcommand{\dthet}{\partial_{\theta}}

\newcommand{\ps}{I}
\newcommand{\psd}{i}
\newcommand{\dmu}{\partial_{\mu}}

\newcommand{\clr}{{\mathsf{R}}}

\newcommand{\Kp}{\bar{K}}
\newcommand{\Gp}{\bar{G}}
\newcommand{\Sp}{\bar{S}}
\newcommand{\Jp}{\bar{J}}

\begin{document}

\begin{center}

{\bf\large NEW TECHNIQUES FOR ANALYSING AXISYMMETRIC } \\
{\bf\large GRAVITATIONAL SYSTEMS.  1 VACUUM FIELDS} \\

\vspace{0.4cm}
{\large Chris Doran and Anthony Lasenby}

{\it Astrophysics Group, Cavendish Laboratory, Madingley Road,} \\
{\it Cambridge CB3 0HE, UK.}

\vspace{0.4cm}

\today

\vspace{0.4cm}

\begin{abstract}
A new framework for analysing the gravitational fields in a stationary,
axisymmetric configuration is introduced.  The method is used to
construct a complete set of field equations for the vacuum region
outside a rotating source.  These equations are under-determined.
Restricting the Weyl tensor to type~D produces a set of equations
which can be solved, and a range of new techniques are introduced to
simplify the problem.  Imposing the further condition that the
solution is asymptotically flat yields the Kerr solution uniquely.
The implications of this result for the no-hair theorem are discussed.
The techniques developed here have many other applications, which are
described in the conclusions.
\end{abstract}

\end{center}

\section{Introduction}

The field equations of general relativity are remarkable both for
their complexity and for the incredible variety of techniques that can
be applied in generating solutions.  These range from brute force
computation through to advanced techniques based on symmetries and
invariants~\cite{kra-exact,olv-eqv}.  In this paper we introduce a new
method for tackling the gravitational field equations for stationary,
axisymmetric systems.  The main applications of this scheme are to
finding the gravitational fields inside and outside rotating sources.
The method has strong similarities with the Newman--Penrose (NP)
formalism~\cite{kra-exact}, and also with coframe methods for
analysing differential equations.  The essential difference with the
NP formalism is that our approach is based on a real orthonormal
tetrad, as opposed to a complex null tetrad.  The only complex
structure we work with arises naturally in the structure of the
bivector fields generated by the tetrad vectors.

The main advance in the work presented here is that the technique can
be pushed right through to an explicit solution.  This is possible
with the introduction of a number of new methods.  The closest
precursor to these techniques is the work of Held~\cite{held74},
though we believe our current methods offer many improvements.  In
this paper we concentrate on the vacuum equations.  We extract a complete
set of vacuum equations, and find that these are under-determined for
a general, axisymmetric source.  To construct a unique solution we
then impose the additional condition that the Riemann tensor is
type~D, and the solution is asymptotically flat.  We are then able to
show constructively that the Kerr is the unique solution for this
case.  This result does not rely on the existence of a horizon.  The
Carter--Robinson uniqueness theorem can therefore be interpreted as
stating that the formation of a horizon is related to a restriction on
the algebraic form of the Weyl tensor.

The Kerr solution occupies a unique place amongst the known exact
solutions to the Einstein field equations.  Many derivations of the
solution have been found, a number of which are summarised in
Chandrasekhar's classic text~\cite{cha83}.  Few authors (apart from
Chandrasekhar) have claimed that the derivation of the Kerr solution
is simple, and many of the derivations found are highly opaque.
Probably the most mysterious derivation of all is that of the complex
coordinate transformation trick~\cite{new65,inv-rel}.  In this
derivation one starts with the Schwarzschild metric in advanced
Eddington-Finkelstein coordinates, expressed in terms of a complex
null tetrad.  A complex coordinate transformation is then applied to
yield a new metric, which turns out to be that of the Kerr solution.
This is a trick because there is no reason to expect the complex
coordinate transformation to generate a new vacuum solution.  The
justification for this is only revealed at the end of a detailed study
of vacuum metrics of Kerr-Schild type, and is quite
obscure~\cite{sch73}.

The derivation presented here proceeds in a very different way to the
usual `metric' route adopted in general relativity (GR).  In a typical
GR derivation one starts with a metric of suitable symmetry containing
a set of arbitrary scalar functions.  The Einstein equations are then
written as a set of coupled second-order differential equations in the
metric functions.  These equations are notoriously difficult to
analyse, for many different reasons.  One of the difficulties is that
the freedom to perform coordinate transformations must be removed by
restricting the form of the metric (`gauge-fixing').  If this is not
done correctly, one has more unknowns than there are equations for.
The problem is that it is often not clear \textit{ab initio} how best
to perform this gauge fixing.  Ideally one would like the chosen
coordinates to have some simple physical interpretation, but how to
achieve this does not usually emerge until later in the solution
process.

Our method avoids this problem by developing a first-order set of
equations analogous to those obtained in the NP formalism.
Gauge-fixing of the Lorentz group is performed at the level of the
Riemann tensor, which makes it easier to ensure that the gauge choices
are physically sensible and mathematically convenient.  The resulting
equations relate abstract derivatives of the terms in the spin
connection to quadratic combinations of the same quantities.  This
provides a very clear way of expressing and analysing the
non-linearities in the theory.  Once the Riemann tensor has been
found, some natural physical scalar fields start to emerge.
Invariance under diffeomorphisms is then employed to give these
physical fields simple expressions in terms of our chosen coordinates.
This process lifts the status of the coordinates from arbitrary
spacetime functions to local physically-measurable fields.  We then
introduce a series of new techniques which solve directly for the spin
coefficients without computing any of the metric coefficients.  The
latter are then found by straightforward integration.  This is a
considerable advance on previous first-order formulations of GR, and
opens up a number of new solution strategies.  We do not yet possess a
general method for solving all problems in our new formalism, but
there are good reasons to believe that some very general techniques do
exist.  The derivation of the Kerr solution presented here provides a
number of clues to how such general methods might be found.

One feature of our method, which it shares with the NP formalism, is
that the early stages involve simple, repetitive algorithms for
organising terms.  This work is well suited to symbolic algebra
packages such as Maple.  The derivation of the equations is further
simplified by employing the language of Clifford
algebra~\cite{ben-spin,hes-gc,DGL98-grav,chi89}.  This is employed as
a set of algebraic rules for manipulating vectors, rather than via a
concrete matrix representation (such as the Dirac matrices).  In this
sense we are adopting the language of spacetime
algebra~\cite{hes-sta,DGL98-grav}.  This approach to Clifford algebra
is also well suited to implementation in symbolic algebra packages.
The application of spacetime algebra clearly demonstrates the origin
of a natural complex structure generated by the spacetime directed
volume element, or pseudoscalar.  This element squares to minus one,
so unites the complex structure of the 2-spinor approach with duality
transformations in tensor language.  This is a considerable
unification, and provides a number of insights into the complex
structure underlying the Kerr solution.  This also sheds some light on
the origin of the mysterious complex coordinate transformation trick
discovered by Newman and Janis~\cite{new65}.

This paper starts by introducing the variables we use to encode a
stationary axisymmetric gravitating system.  We then introduce a
suitably general ansatz to describe axially-symmetric fields.  The
link between the gauge fields and the metric and Christoffel
connection is explained.  The complete set of vacuum equations is
obtained, and we then restrict to looking for vacuum solutions of
type~D.  Imposing this restriction causes the equations to simplify in
a truly remarkable manner to a set of 8 coupled equations.  These
equations are then simplified by identifying various integrating
factors.  This enables us to make a range of deductions about the
solution without introducing a definite coordinatisation.  Finally, a
concrete expression of the solution is produced, which reproduces the
line element of the Kerr solution in Boyer--Lindquist coordinates.  We
end with a discussion of the insights the present formalism brings to
the no hair theorem for black holes.  This states that the the Kerr
solution is the only possible vacuum axisymmetric solution outside a
horizon, which is now interpreted in terms of a restriction on the
algebraic type of the Weyl tensor at a horizon.

Summation convention and natural units ($c=G=1$) are employed
throughout.  Greek letters are employed for coordinate indices and
Latin for tetrad components.  We work in a space with signature ($1$,
$-1$, $-1$, $-1$).

\section{Field equations for axisymmetric systems}

The equations derived here follow the route developed in an earlier
paper~\cite{DGL98-grav}, though the present derivation is streamlined
and much of the less conventional terminology has been removed.  A
small price for this is that some results in the following derivation
are stated without proof.  Further details can be found
in~\cite{DGL98-grav,DGL-banff}.  Our starting point is the Clifford
algebra of Minkowski spacetime.  The (constant) generators of this are
denoted by $\{\gam_0 \cdots \gam_3\}$ and satisfy
\begin{equation}
\gam_i \gam_j + \gam_j\gam_i = 2\eta_{ij}, \qquad i,j = 0 \ldots 3,
\end{equation}
where our convention is that $\eta_{ij} = \mbox{diag}(1, -1, -1, -1)$.
We suppress any mention of the identity matrix, which is superfluous
for all calculations.  Throughout we use Latin indices for tetrad
frames and components, and Greek for coordinate indices.  At various
points we adopt the language of spacetime algebra and refer to the
Clifford generators as `vectors'.

The symmetric and antisymmetric parts of the Clifford product of two
vectors define the inner and outer products, and are denoted with a
dot and a wedge respectively:
\begin{align}
\gam_i \dt \gam_j &= \half (\gam_i \gam_j + \gam_j \gam_i) =
\eta_{ij} \nn \\
\gam_i \wdg \gam_j &= \half (\gam_i \gam_j - \gam_j \gam_i).
\end{align}
A full (real) basis for the Clifford algebra is provided by
\begin{equation} 
\begin{array}{ccccc}
1 & \{\gam_i\} & \{\gam_i \wdg \gam_j\} & \{\ps \gam_i\} & \ps  \\
\mbox{1 scalar} & \mbox{4 vectors} & \mbox{6 bivectors} & 
\mbox{4 trivectors} & \mbox{1 pseudoscalar}, \\
\mbox{grade 0} & \mbox{grade 1} & \mbox{grade 2} & \mbox{grade 3} &
\mbox{grade 4}
\end{array} 
\label{basis}
\end{equation}
where
\begin{equation} 
\ps = \go\gi\gj\gk.
\end{equation} 
The 4-vector $\ps$ is the highest grade element in the algebra and is
usually called the pseudoscalar, though directed volume element is
perhaps more appropriate.  The pseudoscalar squares to $-1$, 
\begin{equation}
I^2 = -1,
\label{I2}
\end{equation}
and $\ps$ anticommutes with all odd-grade elements and commutes with
even-grade elements.

We next introduce a set of generators appropriate for the study
of axisymmetric fields.  With $(r,\theta,\phi)$ denoting a standard
set of polar coordinates we define
\begin{align}
\gamma_t &= \go \nn \\
\gamma_r &=  \sin\!\theta (\cos\!\phi \, \gi + \sin\!\phi
\, \gj) + \cos\!\theta \, \gk \nn \\
\gamma_\theta  &= \cos\! \theta (\cos\!\phi \, \gi + \sin\!\phi
\, \gj) - \sin\!\theta \, \gk \nn \\
\gamma_\phi &=  -\sin\!\phi \, \gi + \cos\!\phi \, \gj.
\end{align}
These vectors also form an orthonormal basis for Minkowski spacetime.
Henceforth the Latin indices $i=0 \ldots 3$ are assumed to run over
the set $\{ \gamma_t, \gamma_r, \gamma_\theta, \gamma_\phi\}$.  The
reciprocal frame is given by
\begin{equation}
\gamma^t = \gamma_t, \quad 
\gamma^r = -\gamma_r, \quad 
\gamma^\theta = -\gamma_\theta, \quad 
\gamma^\phi = - \gamma_\phi.
\end{equation}

The (Clifford) product of two orthogonal vectors results in a
bivector.  There are six of these in total, and a useful basis for
these is provided by
\begin{align} 
\sigr &=  \gam_r \gam_t \nn \\
\sigth &= \gam_\theta \gam_t \nn \\ 
\sigph &= \gam_\phi \gam_t.
\end{align}
These all have unit square,
\begin{equation}
\sigr\sigr =  \sigth\sigth = \sigph\sigph = 1
\end{equation}
and distinct bivectors anticommute,
\begin{equation}
\sigr\sigth = - \sigth \sigr, \qquad \mbox{\textit{etc}}.
\end{equation}
The three bivectors also satisfy the identity
\begin{equation} 
\sigr \sigth \sigph = \gam_t \gam_r \gam_\theta \gam_\phi =
\go\gi\gj\gk = \ps.
\end{equation}
The bivector basis is completed by the dual bivectors 
\begin{align} 
\ps \sigr &= -\gam_\theta \gam_\phi \nn \\
\ps \sigth &= \gamma_r \gamma_\phi \nn \\
\ps \sigph &= - \gamma_r \gamma_\theta.
\end{align}
The algebra generated by the set $\{\sigr,\sigth,\sigph\}$ is
isomorphic to the Pauli algebra.  This is the reason for the
notation.  It should be remembered, however, that each of the
$\{\sigr,\sigth,\sigph\}$ set anticommutes with $\gam_t$.

The space of bivectors generates the Lie algebra of the group of
Lorentz transformations.  Under a proper orthochronous Lorentz
transformation a vector $a$ is mapped to the vector $a'$ according to
the rule
\begin{equation}
a \mapsto a' = R a \Rrev
\label{rottrf}
\end{equation}
where $R$ is an even-grade Clifford element satisfying
\begin{equation}
R \Rrev = 1.
\end{equation}
The tilde on $\Rrev$ denote the \textit{reverse} operation, which
reverses the order of all products in a general element.  The effect
of this on the grade-$r$ element $M_r$ is
\begin{equation}
\tilde{M}_r = (-1)^{r(r-1)/2} M_r.
\end{equation}
The object $R$ is referred to as a \textit{rotor}.  Rotors form a
group under the Clifford product, which gives a spin-1/2
representation of the proper orthochronous Lorentz group.  All rotors
can be written as $\pm \exp(B/2)$, where $B$ is a bivector.  The
commutator of a bivector and a grade-$r$ object results in a new
object of grade-$r$.  For this operation it is useful to define the
commutator product by
\begin{equation}
M \crs N = \half (MN-NM),
\end{equation}
where $M$ and $N$ are general Clifford elements.  The space of
bivectors is closed under the commutator product, and this generates
the Lie algebra for the rotor group.

The key to our approach is to parameterise the gravitational fields
via a coframe.  We introduce the four vector fields
\begin{align}
g^t &= a_1 \gam^t + \frac{a_2}{r \sin\!\theta} \gam^\phi \nn \\
g^r &= b_1 \gam^r + \frac{b_2}{r} \gam^\theta \nn  \\
g^\theta &= \frac{c_1}{r} \gam^\theta + c_2 \gam^r \nn \\
g^\phi &= \frac{d_1}{r \sin\!\theta} \gam^\phi + d_2 \gam^t,
\label{defg}
\end{align}
where all of the variables $(a_1 \ldots d_2)$ are scalar functions
of $r$ and $\theta$.  The indices on the $\{g^t,g^r,g^\theta,g^\phi\}$
are to be read as coordinate indices, and we denote the set by
$\{g^\mu\}$, $\mu=0\ldots 3$.  The reciprocal frame $\{g_\mu\}$ to the
$\{g^\mu\}$ frame is defined by the equation
\begin{equation}
g_\mu g^\nu + g^\nu g_\mu = 2\delta_\mu^\nu.
\end{equation}

The $g_\mu$ and $\gam^i$ vectors define a tetrad field in the obvious
manner,
\begin{equation}
g_\mu \gam^i + \gam^i g_\mu = 2{h_\mu}^i.
\label{tetrad}
\end{equation}
The metric tensor $g_{\mu\nu}$ is derived from the $g_\mu$ vectors
via
\begin{equation}
g_\mu g_\nu + g_\nu g_\mu = 2
g_{\mu\nu}. 
\label{metric}
\end{equation}
This results in the line element
\begin{align}
ds^2 &= 
\frac{{d_1}^2 - {d_2}^2 r^2 \sin^2\! \theta}{(a_1 d_1-a_2 d_2)^2} dt^2 +
\frac{a_1 d_2 r^2 \sin^2\! \theta - d_1 a_2 }{(a_1 d_1-a_2 d_2)^2} 2
dt \, d\phi - 
\frac{{a_1}^2 r^2 \sin^2\! \theta - {a_2}^2}{(a_1 d_1-a_2 d_2)^2}
d\phi^2 \nn \\
& \quad -
\frac{{c_1}^2 + {c_2}^2 r^2}{(c_1 b_1 - c_2 b_2)^2} dr^2 -
\frac{c_1 b_2 + b_1 c_2 r^2}{(c_1 b_1 - c_2 b_2)^2} 2 dr \, d\theta -
\frac{{b_1}^2 r^2 + {b_2}^2}{(c_1 b_1 - c_2 b_2)^2} d\theta^2.
\label{lineelm}
\end{align}
At first sight this looks unnecessarily complicated.  Various
nonlinearities are introduced, and there are redundant degrees of
freedom in the $g^\mu$ vectors.  But our method works directly with
the $g^\mu$ and not with the line element.  This gives greater
control over the nonlinearities and the extra (gauge) degrees of
freedom can be employed to advantage later in the calculation.

Our main differential operators are the coframe derivatives $L_i$,
\begin{equation}
L_i = \gam_i \dt g^\mu \partial_\mu,
\end{equation}
where we employ the abbreviation
\begin{equation} 
\partial_\mu = \deriv{ }{x^\mu}.
\end{equation}
The coframe derivatives can be viewed as directional derivatives, with
directions determined by the coframe.  Some texts~\cite{olv-eqv}
prefer to use a partial-derivative notation for coframe derivatives.
We have not chosen such a notation because coframe derivatives do not
necessarily commute.  Written out explicitly the $L_i$ operators are
\begin{align}
L_t &=  a_1 \dift + d_2 \dphi \nn \\
L_r &= b_1 \dr + c_2 \dthet \nn \\
L_\theta &= \frac{1}{r} (c_1 \dthet + b_2 \dr) \nn \\
L_\phi &= \frac{1}{r\sin\!\theta}(d_1 \dphi + a_2 \dift).
\end{align}
The coframe derivatives satisfy a set of bracket identities
\begin{equation}
[L_i,L_j] = {c_{ij}}^k L_k.  
\end{equation}
The spin connection is encoded in a set of bivector fields $\om_i$.
These are related to  the ${c_{ij}}^k$ coefficients by
\begin{equation}
{c_{ij}}^k = \gam^k \dt (L_i \gam_j + \om_i \crs \gam_j -
L_i \gam_i -  \om_j \crs \gam_i).
\label{defcs}
\end{equation}
This equation embodies the content of the first structure equation.

Rather than solving this equation directly, our method involves
introducing a general parameterisation for the connection terms.  An
appropriate parameterisation is defined by
\begin{align}
\om_t &= -(T+\ps J) \sigr - (S+\ps K) \sigth - d_2 \gam_1 \gam_2 \nn \\
\om_r &= - (\Sp +\ps \Kp ) I\sigph + c_2 I \sigph \nn \\
\om_\theta &=  - (\Gp +\ps \Jp ) I\sigph + \frac{c_1}{r} I \sigph \nn \\
\om_\phi &= - (H+\ps K) I \sigr + (G+\ps J)I \sigth - \frac{d_1}{r
\sin\!\theta} \gam_1 \gam_2.
\label{om}
\end{align}
This parameterisation introduces a set of 10 scalar functions ($G$,
$\Gp $, $J$, $\Jp $, $S$, $\Sp $, $K$, $\Kp $, $T$, $H$).  Each of
these is a scalar function of $r$ and $\theta$.  The reason for the
labelling scheme will become apparent when the final set of equations
is derived.  Equation~(\ref{defcs}) now produces the 6 bracket relations
\begin{align}
{ } [L_t, L_r] &=  -T L_t -(K+\Kp ) L_\phi &\qquad  
[L_r, L_\theta] &=  -\Sp L_r -\Gp  L_\theta \nn \\
[L_t, L_\theta] &= -SL_t + (J-\Jp ) L_\phi &
[L_r, L_\phi] &= -(K-\Kp ) L_t -G L_\phi \nn \\
[L_t, L_\phi] &= 0 &
[L_\theta,L_\phi] &= (J+\Jp )L_t  -HL_\phi. 
\label{brack}
\end{align}
The Christoffel connection coefficients are recovered from our
various fields through the formula
\begin{equation}
\Gamma^\lambda_{\mu\nu} = g^\lambda \dt (\dmu g_\nu + g_\mu \dt
\gamma^i \, \om_i \crs g_\nu).
\end{equation}
Clearly this cannot be computed until the $g_\mu$ vectors are
explicitly constructed, which is not achieved until the end of the
calculation.

\section{Gauge freedom}

There are two distinct types of gauge freedom present in the setup
described above.  These correspond to the two symmetries on which the
gauge treatment of gravity is based~\cite{DGL98-grav,kib61}.  The
first is invariance under diffeomorphisms.  This can be encoded in
various different ways, but for our present purposes it is simplest to
consider two new scalar coordinates $r'(r,\theta)$ and
$\theta'(r,\theta)$.  Suppose first that we replace $r$ and $\theta$
with $r'$ and $\theta'$ everywhere in the metric of
equation~(\ref{lineelm}).  Clearly such a relabelling does not affect the
solution.  We then re-express the line element back in the original
$(r,\theta)$ coordinate system.  The result is that all of the fields
$(a_1 \ldots d_2)$ have transformed to new fields.  Four of these,
$a_1$, $a_2$, $d_1$, $d_2$ transform according to the simple rule
\begin{equation}
a_1 (r,\theta) \mapsto a_1^\prime(r,\theta) = a_1 (r',\theta').
\label{covtrf}
\end{equation}
Any field transforming in this manner is said to transform as a
covariant scalar under diffeomorphisms.  All of the 10 scalar
functions introduced in the $\om_i$ of equation~(\ref{om}) behave as
covariant scalars.

The remaining variables in the $g^\mu$ vectors, $b_1$, $b_2$,
$c_1$, $c_2$, pick up derivatives of the coordinate transformation
under the diffeomorphism described above.  The transformed variables
are given by
\begin{equation}
\begin{pmatrix}
	b_1^\prime (r,\theta) & c_2^\prime (r,\theta) \\
	b_2^\prime (r,\theta) & c_1^\prime (r,\theta)/r
\end{pmatrix}
= 
\begin{pmatrix}
	b_1 (r',\theta') & c_2 (r',\theta') \\
	b_2 (r',\theta') & c_1 (r',\theta')/r' 
\end{pmatrix}
\begin{pmatrix}
	\partial_{r'} r & \partial_{r'} \theta \\
	\partial_{\theta'} r & \partial_{\theta'}\theta 
\end{pmatrix}.
\label{displ}
\end{equation}
It is simplest for our purposes to view the eight transformed
variables $a_1^\prime \ldots d_2^\prime$ as a new set of functions
obtained from the original set by a gauge transformation.  As such,
the two sets are physically indistinguishable, and this gauge freedom
must be fixed before we can write down a unique solution.  The
standard, metric-based approach to solving the Einstein equations
usually attempts to fix this gauge freedom at the outset by
restricting the terms in the metric.  For example we could write the
line element in the form~\cite{cha83}
\begin{equation}
ds^2 = \et{2\alpha} \, dt^2 - \et{2\psi} (d\phi - \om \, dt)^2 -
\et{2\mu_2} \, dr^2 - \et{2\mu_3} \, d\theta^2.
\end{equation}
This leaves us with a set of 5 variables, with the freedom to fix the
relation between $\mu_2$ and $\mu_3$ by a further coordinate
transformation.  Our strategy (in keeping with that of the NP
formalism) is to leave this gauge unfixed until much later in the
calculation.  This is possible because the coframe derivatives
transform as
\begin{equation}
L_r \mapsto L_r^\prime = b_1(r',\theta') \deriv{ }{r'} +
c_2(r',\theta') \deriv{ }{\theta'}
\end{equation}
with a similar result holding for $L_\theta$.  Provided we formulate
all equations in terms of coframe derivatives of covariant scalars,
the entire structure will transform covariantly under
diffeomorphisms.  That is, the gauge freedom can be ignored until
later in the problem.  We then find that certain physical fields
emerge and it is sensible to equate these with combinations of the
chosen coordinates.  This is the point at which the diffeomorphism
gauge freedom is fixed.  Working in this manner ensures that the
chosen coordinates have a direct physical interpretation.

The second gauge freedom is that of applying a local Lorentz rotation
to the $g^\mu$ vectors.  This is the gravitational analogy of a
phase transformation in quantum theory.  Under a Lorentz rotation the
$g^\mu$ vectors transform according to
\begin{equation}
g^\mu \mapsto  R g^\mu \Rrev
\label{rttrf}
\end{equation}
where $R$ is a local rotor.  The reciprocal vectors $g_\mu$ transform the
same way, and the metric derived by equation~(\ref{metric}) is therefore
invariant.  Clearly this freedom is a gauge symmetry, as it does not
change any physical quantity.  This symmetry is removed if one works
directly with terms in the metric, but our approach is to keep the
symmetry explicit and use it to simplify our equations.  The $\om_i$
bivectors represent the connection for the gauge symmetry of
equation~(\ref{rttrf}).  The combination
\begin{equation}
D_\mu = \partial_\mu + g_\mu \dt \gamma^i \, \om_i \times
\end{equation}
defines the covariant derivative for objects transforming in the
manner of equation~(\ref{rttrf}).  For our axisymmetric setup the degrees
of freedom in the Lorentz gauge are reduced from six to two.  These are
a rotation in the $I\sigph$ plane and a boost in the $\sigph$
direction.  The rotors describing these can be combined to form the
single rotor
\begin{equation}
R = \exp(\alpha \sigph /2) \exp(\beta \ps\sigph /2) = \exp(w \ps
\sigph /2), 
\label{rott}
\end{equation}
where $w=\beta-I \alpha$ is an arbitrary function of ($r$, $\theta$).
The freedom to apply this rotor will be used to simplify the Riemann
tensor.

One final (non-gauge) freedom present is that the equations are
unchanged if all variables are scaled by a constant amount, $a_1
\mapsto c a_1$, $G \mapsto cG$, \textit{etc}.  This freedom is often
employed to fix the asymptotic behaviour so that the $g_\mu$ vectors
map to a flat space coordinate frame at spatial infinity.

\section{The Riemann tensor and vacuum fields}
\label{S-Riem}

The essential covariant object to construct is the Riemann tensor.
The information contained in this is compactly summarised in a set of
six bivectors.  These are defined by
\begin{equation}
\clr_{ij} = L_i \om_j - L_j \om_i + \om_i \crs
\om_j - {c_{ij}}^k \om_k.
\end{equation}
Calculating each of the terms $\clr_{ij}$ is simply a matter of
organisation, and is well suited to a symbolic algebra package.  A
useful set of intermediate results is provided by the identities
\begin{align}
L_t \sigr + \om_t \crs \sigr &= (S+\ps K) \ps\sigph & \qquad 
L_t \sigth + \om_t \crs \sigth &= -(T+\ps J)\ps\sigph \nn \\ 
L_r\sigr + \om_r \crs \sigr &= (\Sp +\ps \Kp ) \sigth &
L_r \sigth + \om_r \crs \sigth &= -(\Sp +\ps \Kp )\sigr \nn \\
L_\theta\sigr + \om_\theta \crs \sigr &= (\Gp +\ps \Jp ) \sigth &
L_\theta \sigth + \om_\theta \crs \sigth &= -(\Gp +\ps \Jp ) \sigr \nn \\ 
L_\phi\sigr + \om_\phi \crs \sigr &= (G+\ps J)\sigph &
L_\phi \sigth + \om_\phi \crs  \sigth &= (H+\ps K)\sigph.
\label{sigds}
\end{align}
On substituting the form for the $\om_i$ into the definition of
$\clr_{ij}$ we now obtain the Riemann tensor in terms of abstract
derivatives as
\begin{align}
\clr_{rt} 
&= \Bigl( -L_r(T+\ps J) + (S+\ps K)(\Sp +\ps \Kp ) +T(T+\ps J) +\ps
(K+\Kp )(H+\ps K) \Bigr) \sigr \nn \\ 
&\quad + \Bigl( -L_r(S+\ps K) -(\Sp +\ps \Kp )(T+\ps J) +T(S+\ps K) -\ps
(K+\Kp )(G+\ps J) \Bigr) \sigth \nn \\  
\clr_{\theta t} 
&= \Bigl( -L_\theta(S+\ps K) -(\Gp +\ps \Jp ) (T+\ps J) +S(S+\ps K) + \ps
(J-\Jp )(G+\ps J) \bigr) \sigth \nn \\ 
&\quad + \Bigl( -L_\theta (T+\ps J) +(\Gp +\ps \Jp ) (S+\ps K) +S
(T+\ps J) -\ps  (J-\Jp )(H+\ps K) \Bigr) \sigr \nn \\
\clr_{\phi t} 
& = - \Bigl( (G+\ps J)(T+\ps J) + (S+\ps K)(H+\ps K) \Bigr)  \sigph \nn \\
\clr_{r\theta} 
&= - \Bigl( L_r(\Gp +\ps \Jp ) - L_\theta (\Sp +\ps \Kp ) +\Gp (\Gp +\ps \Jp )
+\Sp (\Sp +\ps \Kp ) \Bigr) \ps \sigph \nn \\
\clr_{r\phi}
&= \Bigl( L_r(G+\ps J) - (\Sp +\ps \Kp ) (H+\ps K) +G(G+\ps J) +\ps
(K-\Kp )(S+\ps K) \Bigr) \ps \sigth \nn \\  
& \quad - \Bigl( L_r(H+\ps K) +(\Sp +\ps \Kp ) (G+\ps J) +G(H+\ps K) -\ps
(K-\Kp )(T+\ps J) \Bigr)  \ps\sigr \nn \\
\clr_{\theta\phi}
&= - \Bigl(L_\theta (H+\ps K) +(\Gp +\ps \Jp ) (G+\ps J) + H(H+\ps K) + \ps
(J+\Jp )(T+\ps J) \Bigr) \ps\sigr \nn \\ 
&\quad + \Bigl( L_\theta (G+\ps J) -(\Gp +\ps \Jp )(H+\ps K) +H (G+\ps J)
-\ps (J+\Jp )(S+\ps K) \Bigr)  \ps\sigth. 
\label{Riem}
\end{align} 
The standard coordinate expression of the Riemann tensor is obtained
from the preceding bivectors by forming the contraction
\begin{equation}
{R^\mu}_{\nu \rho \sig} = (g^\mu \wdg g_\nu) \dt \clr_{ij} \, \gam^i \dt
g_\sig \, \gam^j \dt g_\rho.
\end{equation}
As with the Christoffel connection, these coefficients cannot be
written down until an explicit form is obtained for the $g^\mu$
vectors.

By working with a general form for the spin connection we have no
guarantee that the Riemann tensor satisfies all of its required
symmetries.  The first of these to consider arises from the assumption
that there is no torsion present.  In this case the Riemann tensor
satisfies
\begin{equation}
\gam^i \wdg \clr_{ij} = \half(\gam^i\clr_{ij}+ \clr_{ij} \gam^i ) = 0.
\label{Riemsym}
\end{equation}
This says that a trivector vanished for each value of the index $j$,
so corresponds to a set of 16 constraints.  The Riemann
tensor is encoded in six bivector-valued functions, giving a total of
36 scalar degrees of freedom.  Equation~(\ref{Riemsym}) reduces this to
the familiar 20 degrees of freedom of a general Riemann tensor.

In this paper we are interested in vacuum solutions to the field
equations.  For these the contraction of the Riemann tensor must also
vanish, 
\begin{equation}
\gam^i \crs \clr_{ij} = \half(\gam^i\clr_{ij} - \clr_{ij} \gam^i ) = 0.
\label{vac}
\end{equation}
This can be combined with equation~(\ref{Riemsym}) to obtain
\begin{equation}
\gam^i  \clr_{ij} = 0. 
\label{Fullcnd}
\end{equation}
This equation is satisfied by all Weyl tensors. Equation~(\ref{vac})
reduces the number of degrees of freedom in the vacuum Riemann tensor
from 20 to 10 --- the expected number for vacuum solutions.  The
compact combination of the symmetry and contraction information in
equation~\ref{Fullcnd} is unique to the Clifford algebra formulation.

Before proceeding, it is useful to adopt a slightly different notation
for the Riemann tensor.  We define a linear map from bivectors to
bivectors by writing
\begin{equation}
\clr(B) = \half B \dt ( \gam^j \wdg \gam^i) \, \clr_{ij} .
\end{equation}
Equation~(\ref{Riem}) then shows that the Riemann tensor has the general form
\begin{align}
\clr(\sigr) &= \alpha_1 \sigr + \beta_1 \sigth & 
\clr(\ps\sigr) &= \alpha_4 \ps\sigr + \beta_4 \ps\sigth \nn \\
\clr(\sigth) &= \alpha_2 \sigth + \beta_2 \sigr &
\clr(\ps\sigth) &= \alpha_5 \ps\sigth + \beta_5 \ps\sigr \nn \\
\clr(\sigph) &= \alpha_3 \sigph &
\clr(\ps\sigph) &= \alpha_6 \ps\sigph,
\end{align}
where each of the $\alpha_i$ and $\beta_i$ are scalar $+$ pseudoscalar
combinations.  We can now clearly see how a complex structure emerges
based on the spacetime pseudoscalar.  On writing out of the four
equations~(\ref{Fullcnd}) in full, and pre-multiplying each equation by
$\gam_j$, we obtain the four equations
\begin{align}
\sigr \clr (\sigr) + \sigth \clr (\sigth) + \sigph \clr (\sigph) 
&= 0 \nn \\
\sigr \clr (\sigr) - I\sigth \clr (I\sigth) - I\sigph \clr (I\sigph) 
&= 0 \nn \\
-I\sigr \clr (I\sigr) + \sigth \clr (\sigth) -I \sigph \clr (I\sigph) 
&= 0 \nn \\
-I\sigr \clr (I\sigr) -I \sigth \clr (I\sigth) + \sigph \clr (\sigph) 
&= 0 .
\end{align}
Summing the final three equations and employing the first we obtain 
\begin{equation}
I\sigr \clr (I\sigr) + I \sigth \clr (I\sigth) + I\sigph \clr (I\sigph) 
= 0.
\label{Wsym}
\end{equation}
Substituting this back into the final three equations we obtain the
simple relation
\begin{equation}
\clr(I\sig_i) = I \clr(\sig_i), \qquad (i=r,\theta,\phi).
\label{dual}
\end{equation}
This equation shows that the vacuum Riemann tensor (and the Weyl
tensor in general) is linear on the pseudoscalar.  This is the origin
of the complex structure often employed in analysing the Weyl tensor.
Equation~(\ref{dual}) then sets 
\begin{equation} 
\alpha_1 =\alpha_4, \quad \alpha_2=\alpha_5, \quad \alpha_3=\alpha_6,
\quad \beta_1=\beta_4 \quad \beta_2 = \beta_5.
\end{equation} 
Equation~(\ref{Wsym}) says that, viewed as a complex linear function,
the vacuum Riemann tensor is symmetric and traceless.  The most
general form of $\clr(B)$ allowed for vacuum axisymmetric solutions
therefore has the form
\begin{align}
\clr(\sigr) &= \alpha_1 \sigr + \beta \sigth \nn \\
\clr(\sigth) &= \alpha_2 \sigth + \beta \sigr \nn \\
\clr(\sigph) &= - (\alpha_1 + \alpha_2) \sigph .
\label{genvac}
\end{align}

We can simplify equation~(\ref{genvac}) further by recalling the Lorentz gauge
freedom present in our setup.  Under a Lorentz gauge transformation
parameterised by the rotor $R$ the Riemann tensor transforms according
to  
\begin{equation}
\clr(B) \mapsto \clr'(B) = R\clr(\Rrev B R) \Rrev.
\end{equation}
In our axisymmetric setup the rotor $R$ is restricted to the form of
equation~(\ref{rott}).  Under this transformation we find that $\beta$
transforms to
\begin{equation} 
\beta' = \cos (2w) \beta - \half (\alpha_1-\alpha_2)
\sin (2w) 
\end{equation}
where the scalar $+$ pseudoscalar combination $w$ is treated as a
complex number.  We can therefore eliminate the off-diagonal term from
$\clr(B)$ by setting $\tan(w) = 2\beta/(\alpha_1-\alpha_2)$.  This
fixes $w$ and so removes the rotational degrees of freedom.  The
conclusion is that the Riemann tensor for the vacuum outside an
axisymmetric source can be written
\begin{align}
\clr(\sigr) &= \alpha_1 \sigr \nn \\
\clr(\sigth) &= \alpha_2 \sigth \nn \\
\clr(\sigph) &= - (\alpha_1 + \alpha_2) \sigph .
\label{genvac2}
\end{align}
This can be compared with equation~(\ref{Riem}) to obtain a set of 20
equations, which naturally couple into 10 scalar $+$ pseudoscalar
combinations. Four of these equations return explicit formulae for the
action of $L_r$ and $L_\theta$ on $G$, $T$, $J$, $S$, $H$, $K$,
\begin{align}
L_r(S+\ps K) &= -(\Sp +\ps \Kp )(T+\ps J) +T(S+\ps K) -\ps
(K+\Kp )(G+\ps J) \nn \\
L_r(H+\ps K) &= -(\Sp +\ps \Kp ) (G+\ps J) - G(H+\ps K) +\ps
(K-\Kp )(T+\ps J) \nn \\
L_\theta (G+\ps J) &= (\Gp +\ps \Jp )(H+\ps K) -H (G+\ps J)
+\ps (J+\Jp )(S+\ps K) \nn \\
L_\theta (T+\ps J)  &= (\Gp +\ps \Jp ) (S+\ps K) +S
(T+\ps J) -\ps  (J-\Jp )(H+\ps K).
\label{feqn1}
\end{align}
A further four equations return the remaining coframe derivatives of
$G$, $T$, $J$, $S$, $H$, $K$, but also include terms from the Riemann
tensor
\begin{align}
L_r(G+\ps J) &= (\Sp +\ps \Kp ) (H+\ps K) - G(G+\ps J) - \ps
(K-\Kp )(S+\ps K)+\alp_2 \nn \\
L_r(T+\ps J) &= (S+\ps K)(\Sp +\ps \Kp ) +T(T+\ps J) +\ps
(K+\Kp )(H+\ps K) - \alp_1 \nn \\
L_\theta(S+\ps K) &= -(\Gp +\ps \Jp ) (T+\ps J) +S(S+\ps K) + \ps
(J-\Jp )(G+\ps J) - \alp_2 \nn \\
L_\theta (H+\ps K) &= -(\Gp +\ps \Jp ) (G+\ps J) - H(H+\ps K) - \ps
(J+\Jp )(T+\ps J) + \alp_1.
\label{feqn2}
\end{align}
The final equations give
\begin{equation}
\alp_1 + \alp_2 = (G+\ps J)(T+\ps J) + (S+\ps K)(H+\ps K),
\label{defalp}
\end{equation}
and
\begin{equation}
L_r(\Gp +\ps \Jp ) - L_\theta (\Sp +\ps \Kp ) + \Gp (\Gp +\ps \Jp )
+\Sp (\Sp +\ps \Kp ) = \alp_1 + \alp_2.
\label{remeqn}
\end{equation}
So far we are some way short of a fully determined system.  The next
step is to impose the Bianchi identities.

\section{The bracket structure and Bianchi identities}

The Lie bracket structure of equation~(\ref{brack}) gives rise to
a series of higher order constraints.  This information is summarised
in the Bianchi identities.  For vacuum solutions the Bianchi
identities take the simple form~\cite{DGL98-grav}
\begin{equation}
\gam^i \Bigl( L_i (\clr(B)) - \clr(L_i B) + \om_i \crs \clr(B) -
\clr(\om_i \crs B) \Bigr) = 0,
\label{Bchi}
\end{equation}
which holds for any bivector $B$.  The linearity on $I$ and the
properties of the Riemann tensor imply that we only obtain new
information for $B=\sigr$ and $\sigph$.  At this point it is useful to
set 
\begin{align}
\alp &= \alp_1 + \alp_2 \nn \\
\del &= \alp_1 + 2 \alp_2.
\end{align}
The reason for this choice is that $\del=0$ corresponds to a type~D
solution.  Applying equation~(\ref{Bchi}) to $B=\sigph$ yields the pair of
equations
\begin{align}
L_r \alp &= -(3 \alp - \del)(G+\ps J) + \del(T+ \ps J) \nn \\
L_\theta \alp &= (3 \alp - \del)(S+\ps K) - \del (H + \ps K).
\label{diffalpha}
\end{align}
These are entirely consistent with equations~(\ref{feqn1}), (\ref{feqn2})
and~(\ref{defalp}) so contain no new information.  Applied to $B=\sigr$,
however, the Bianchi identity does provide two new equations, 
\begin{align}
L_r \del &= - (3 \alp - \del)(G+\ps J) + 2 \del (T+\ps J) +
(3\alp-2\del) (\Gp + \ps \Jp) \nn \\
L_\theta \del &= (3 \alp - \del)(S+\ps K) - 2 \del (H+\ps K) - 
(3\alp-2\del) (\Sp + \ps \Kp).
\label{delbrk}
\end{align}

We now have expressions for many of the coframe derivatives of the
main physical variables.  These derivatives must all be consistent
with the bracket structure, which reduces to the single identity
\begin{equation}
[L_r,L_\theta] = -\Sp L_r - \Gp L_\theta.
\label{bktidn}
\end{equation}
This identity already holds for all of the unbarred fields.  The
identity can also be applied to $\del$ and yields one further
equation,
\begin{multline}
(3 \alp - 2 \del) \Bigl( L_r (\Sp+\ps \Kp) + L_\theta (\Gp+\ps \Jp) +
\Gp(\Sp+\ps\Kp) -\Sp(\Gp+\ps \Jp) \Bigr) = \\
3 \alp \Bigl( (S+H+2\ps K)( G+\ps J -\Gp -\ps \Jp) - (G+T+2\ps
J)(S+\ps K - \Sp -\ps \Kp ) \Bigr) .
\label{xtreqn}
\end{multline}
But this is as far as the equations can be developed without further
physical information.  There are no further expressions on which to
evaluate the bracket identity, and we do not have a complete set of
coframe derivatives.  That is, the vacuum structure is currently
under-determined.  This is to be expected as there is no unique vacuum
solution outside a rotating star.  To obtain a unique solution we must
either impose suitable boundary conditions, or make a further
restriction on the form of the Riemann tensor.  For example, we could
set $J$, $\Jp$, $K$ and $\Kp$ to zero.  The effect of this is to set
$\alp$ and $\delta$ to real variables, so that the equations reduce
dramatically.  The equations then describe the fields outside a
static, axisymmetric source.  That is, a non-rotating lump of matter
which axial symmetry.  This setup was first discussed by Weyl in
1917~\cite{wey17,kra-exact}.  But before we embark on solving the
general equations for a particular system, we give a comparison of our
equations with the NP formalism.

\section{The Newman--Penrose formalism and complex structures}

In the present scheme the field equations for axisymmetric vacuum
fields are summarised by equations~(\ref{feqn1}), (\ref{feqn2}), (\ref{defalp}),
(\ref{remeqn}), (\ref{delbrk}) and~(\ref{xtreqn}), together with the bracket
identity~(\ref{bktidn}).  There is clearly a close analogy between these
equations and the NP formalism.  The first point to note is that all
equations now consist of scalar $+$ pseudoscalar combinations.  As all
other algebraic elements have been removed, the only remaining effect
of the Clifford pseudoscalar $I$ is to provide a complex structure
through the result~(\ref{I2}) that $I^2=-1$.  We can therefore
systematically replace $I$ by the unit imaginary $i$,
\begin{equation}
I \mapsto i.
\end{equation}
This is quite helpful typographically, as it distinguishes field
variables from the the (constant) pseudoscalar.  We therefore employ
this device at various points.  It must be remembered, however, that
when we construct the full solution in terms of the $\om_i$, the
imaginary $i$ must be replaced by the pseudoscalar.

Clearly, part of the complex structure in the NP formalism has its
geometric origin in the properties of the spacetime pseudoscalar.
This unites the complex structure with spacetime duality.  But the
starting point for the NP formalism is a complex null tetrad.  We
define the four (Minkowski) null vectors
\begin{align}
l &= \frac{1}{\sqrt{2}}(\gam_t + \gam_\phi) & 
m &=  \frac{1}{\sqrt{2}}(\gam_r + i \gam_\theta) \nn \\
n &= \frac{1}{\sqrt{2}}(\gam_t - \gam_\phi) & 
\bar{m} &=  \frac{1}{\sqrt{2}}(\gam_r - i \gam_\theta) .
\label{nullt}
\end{align}
If these vectors replace the $\{\gam_i\}$ in equation~(\ref{tetrad}) the
result is a complex null tetrad.  The differential operators defined
by the null tetrad are, following the conventions of Kramer
\textit{et al.}~\cite{kra-exact}, 
\begin{align}
D &= \frac{1}{\sqrt{2}} (L_t + L_\phi) &
\delta &= \frac{1}{\sqrt{2}} (L_r + i L_\theta) \nn \\
\Delta &= \frac{1}{\sqrt{2}} (L_t - L_\phi) &
\delta^\ast &= \frac{1}{\sqrt{2}} (L_r - i L_\theta).
\end{align}
One can form bracket identities on these derivatives to obtain each of
the spin coefficients in terms of the fields in the $\om_i$.  For
example, we find that
\begin{equation}
\, [\Delta,D] = (\gam + \gam^\ast) D + (\eps+\eps^\ast) \Delta -
(\tau^\ast+\pi) \delta - (\tau+\pi^\ast) \delta^\ast = [L_t,L_\phi] = 0,
\end{equation}
where $\gamma$, $\epsilon$, $\tau$ and $\pi$ are spin coefficients.
Continuing in this manner we obtain expressions for each of the spin
coefficients.  These include the relations 
\begin{equation}
\tau = \frac{1}{2\sqrt{2}} \bigl( T-G + i(S-H) \bigr)
\end{equation}
and
\begin{equation}
\nu = - \frac{1}{2\sqrt{2}} \bigl( G+ T + 2K + i(2J -S -H) \bigr).
\end{equation}
There are two problems with the NP formalism.  The first is that the
key variables (in the spin coefficients) are inappropriate
combinations of the natural variables in the $\om_i$.  The second is
that there are two distinct complex structures at work.  One is
geometric, and has its origin in the pseudoscalar; the other is purely
formal and is inserted by hand in the null tetrad.  If both of these
are represented by the same unit imaginary $i$ then the true geometric
structure is lost.  This is a fundamental flaw, which compromises our
ability to solve the equations.

As an aside, it is possible to formulate the null tetrad~(\ref{nullt}) in
terms of a real Clifford algebra.  In this case the entire complex
structure is provided by the pseudoscalar.  A consequence of this is
that the null tetrad consists of vector and trivector
combinations~\cite{DGL-polspin}.  This is appropriate for some
encodings of supersymmetry, but is not suitable for analysing the
gravitational equations, where the null tetrad must consist of complex
vectors.

\section{Type D vacuum solutions}

In order to take our method on to a solution we need to impose some
extra constraints on the vacuum fields.  Ideally one would like to
impose boundary conditions on some fixed 2-surface representing the
edge of a rotating source, and then propagate the fields out into the
exterior region.  Here we follow a slightly different approach and
restrict the algebraic form of the Riemann tensor to type~D.  The
relationship between this restriction and the existence of a horizon
is described in section~(\ref{S-Unique}).

Viewed as a complex linear function the eigenvalues of the Riemann
tensor~(\ref{genvac2}) are $\alp_1$, $\alp_2$ and $-(\alp_1 + \alp_2)$.
This reduces to a type~D tensor if two of the eigenvalues are equal.
One way of achieving this is to set $\alp_1=\alp_2$. But for this case
the equation structure collapses to a simple system which is not
asymptotically flat.  The remaining two possibilities are gauge
equivalent, so we can restrict to type~D by setting
\begin{equation}
\alp_2 = - \alp_1 - \alp_2.
\end{equation}
It follows immediately that $\del=0$.   The Riemann tensor now
reduces to
\begin{align}
\clr(\sigr) &= 2 \alpha \sigr \nn \\
\clr(\sigth) &= - \alpha \sigth \nn \\
\clr(\sigph) &= - \alpha \sigph ,
\end{align}
where
\begin{equation}
\alpha = (G+\ps J)(T+\ps J) + (S+\ps K)(H+\ps K).
\label{alpha}
\end{equation}
The Riemann tensor can be encoded neatly in the single expression
\begin{equation}
\clr(B) = \half \alpha(B+3 \sig_r B \sig_r),
\label{riem3}
\end{equation}
an expression which is unique to the Clifford algebra formulation.

Returning to equation~(\ref{delbrk}) and setting $\del=0$ we see that we
must now have
\begin{align} 
\Gp +\ps \Jp & = G+\ps J \nn \\
\Sp +\ps \Kp &= S+\ps K.
\end{align} 
This simplification for type~D fields explains our notation using
barred and unbarred variables.  With this simplification we find that
the derivative relations~(\ref{feqn1}) and~(\ref{feqn2}) collapse to give
\begin{align}
L_r(G+\psd J) &= -(G+\psd J)^2 - T(G+\psd J)\nn \\
L_r(T+\psd J) &= (S+\psd K)^2 -\bigl(2(G+\psd J) -T\bigr)(T+\psd J) -2S(H+\psd
K) \nn \\ 
L_r(S+\psd K) &= -\psd J(S+\psd K) - 2\psd K(G+\psd J) \nn \\
L_r(H+\psd K) &= -(G+\psd J)(S+\psd K) -G(H+\psd K) 
\label{reqns}
\end{align}
and
\begin{align}
L_\theta(S+\psd K) &= (S+\psd K)^2 +H(S+\psd K) \nn \\
L_\theta(H+\psd K) &= -(G+\psd J)^2 + \bigl( 2(S+\psd K) -H
\bigr)(H+\psd K) +2G(T+\psd J) \nn \\
L_\theta(G+\psd J) &= \psd K(G+\psd J) +2\psd J(S+\psd K) \nn \\
L_\theta(T+\psd J) &= (G+\psd J)(S+\psd K) +S(T+\psd J). 
\label{theqns}
\end{align}
We have here adopted the convention of representing the pseudoscalar
$I$ with the unit imaginary $i$.  This is a useful device when
analysing this type of equation structure with a symbolic algebra
package.  The above equations are all consistent with the bracket
structure, which now reduces to the single identity
\begin{equation}
[L_r,L_\theta] = -S L_r - G L_\theta.
\label{newbrckt}
\end{equation}
This set of equations is now complete --- we have explicit forms for
the intrinsic derivatives of all of our variables, these are all
consistent with the bracket structure, and the full Bianchi identities
are all satisfied.  Obtaining a set of `intrinsic' equations such as
these is a key step in our method.

The equations~(\ref{reqns}) and~(\ref{theqns}) exhibit a remarkable symmetry.
The $L_r$ and $L_\theta$ derivatives can be obtained from each other
through the interchange
\begin{align}
\qquad \qquad G &\lra S &  J &\lra K \qquad \qquad \nn \\
T &\lra H & L_r &\lra - L_\theta. 
\label{conjug}
\end{align} 
The origin of this symmetry is explained in Section~(\ref{ConjSols}),
where it is related to the conjugacy transformation discussed by
Chandrasekhar~\cite{cha83}.

\section{The Schwarzschild solution}

Before proceeding to the Kerr solution, it is helpful to look at how
the Schwarzschild solution solution fits into our scheme.  The
restriction to spherical symmetry sets all of $a_2$, $b_2$, $c_2$ and
$d_2$ to zero, and we also require that $d_1=c_1$.  The remaining
functions, $a_1$, $b_1$ and $c_1$ are all functions of $r$ only.  The
bracket relations now tell us that the only remaining functions in
$\om_i$ are $T$, $G=\Gp $ and $H$, with $H$ given by
\begin{equation}
H = \frac{c_1 \cos \! \theta}{r \sin \!\theta}.
\end{equation}
We are therefore left with the pair of equations
\begin{align}
L_r G &= -G^2-GT \nn \\
L_r T &= -2GT+T^2,
\label{ss2}
\end{align}
where $L_r=b_1 \dr$.  Dividing these, and solving the resulting
homogeneous differential equation, we obtain
\begin{equation}
GT = \lam_0 (G^2-2GT)^{3/2},
\end{equation}
where $\lam_0$ is an arbitrary constant.  A significant point is that
this relation is derived without fixing the diffeomorphism gauge.  The
relationship is an intrinsic feature of the solution.

The Riemann invariant $\alpha$ is now simply $GT$, and satisfies
\begin{equation}
L_r \alpha = -3G\alpha.
\label{sz3}
\end{equation}
The equation for $L_\theta H$ reduces to the algebraic
expression
\begin{equation}
\frac{c_1^2}{r^2} = G^2 - 2 GT.
\label{ss-i1}
\end{equation}
If follows that
\begin{equation}
\alp = GT = \lam_0 \frac{{c_1}^3}{r^3}.
\end{equation}
We retain the diffeomorphism freedom to perform an $r$-dependent
transformation, which we can employ to fix the functional dependence
of one of our variables.  The obvious way to employ this freedom now
is to set $c_1=1$.  This ensures that $r$ defines the proper
circumference of a sphere in the standard Euclidean manner.  The gauge
choice also ensures that the tidal force, controlled by $\alp$, falls
off as $r^{-3}$, which agrees with the Newtonian result.  It is then
clear, from comparison with the Newtonian result, that $\lam_0$ can be
identified with the mass, so
\begin{equation}
\alpha = \frac{M}{r^3}.
\end{equation}
This ensures that the status of $r$ is lifted from being an arbitrary
coordinate to a physical field controlling the magnitude of the tidal
force.

Equation~(\ref{sz3}) now tell us that $G$ is given by  
\begin{equation}
G=\frac{b_1}{r}
\end{equation}
and equation~(\ref{ss2}) becomes
\begin{equation}
b_1 \deriv{ }{r} \left( \frac{b_1}{r} \right) + \frac{{b_1}^2}{r^2} =
\frac{M}{r^3}. 
\label{swzg1}
\end{equation}
This integrates to give
\begin{equation}
{b_1}^2 = c_0 - 2M/r
\end{equation}
where $c_0$ is the arbitrary constant of integration.  A combination
of a further constant rescaling of $r$ and a global scale change can
be used to set $c_0=1$, so that $g_r$ tends to a flat space polar
coordinate vector at large $r$.

Finally, we recover $a_1$ from the $[L_t,L_r]$ bracket relation, which
yields
\begin{equation}
L_r a_1 = Ta_1.
\end{equation}
Since $L_r b_1=-Tb_1$, it follows that $a_1 b_1$ must be constant.
This constant can be gauged to $1$ by a constant rescaling of the time
coordinate $t$, so we have
\begin{equation}
a_1 = (1-2M/r)^{-1/2}.
\end{equation}
This solves for all of the terms in $\{g^\mu\}$, and calculating the
associated metric recovers the Schwarzschild line element.

This derivation illustrates a number of the key features which
reappear in the derivation of the Kerr solution.  The intrinsic
differential structure can be taken almost completely through to
solution without ever introducing any coordinatisation.  Coordinates
are only fixed in the final step, when one solves for the metric
coefficients.  For the Schwarzschild case this results in
equation~(\ref{swzg1}), which is solved directly be integration and could
have been computed numerically had a simple analytic solution not been
available.  All of the remaining terms are found algebraically, or by
integration alone.  Another feature is the employment of any residual
gauge freedom to set the asymptotic conditions so that the $\{g_\mu\}$
vectors map to a flat space coordinate frame as $r\mapsto \infty$.
Finally, the form of the solution shows that the fields are only valid
for $r>2M$.  Our initial ansatz~(\ref{defg}) is not sufficiently general
to describe a global solution when a horizon is present.  This problem
is only resolved by introducing further terms to allow for a $t$--$r$
coupling~\cite{DGL98-grav}.

\section{Simplifying the Kerr equations}

We are now in a position to complete the derivation of the Kerr
solution.  Our approach is to construct a solution which is
asymptotically flat at large $r$, for all angles.  This condition
turns out to be sufficient to construct a unique solution, which is
then easily seen to be the Kerr solution.

The first step in solving the equation structure of~(\ref{reqns})
and~(\ref{theqns}) is the identification of a number of integrating
factors.  If a pair of variables $A$ and $B$ satisfy the equation
\begin{equation}
L_\theta A - L_r B = GB + SA
\label{intfaccon}
\end{equation}
then an integrating factor $x$ exists defined (up to an arbitrary
magnitude) by
\begin{equation}
L_r x = Ax, \qquad L_\theta x = Bx.
\end{equation}
Equation~(\ref{intfaccon}) ensures that the definition of the integrating
factor is consistent with the bracket structure~(\ref{newbrckt}).  This
means that, when a coordinatisation is chosen, the equations for the
partial derivatives of $x$ are consistent, and $x$ can then be
computed by straightforward integration.

The first integrating factor is constructed from $\alpha$.  The factor
of 3 in equation~(\ref{diffalpha}) prompts us to define
\begin{equation}
Z = Z_0 \alpha^{-1/3} ,
\end{equation}
where $Z_0$ is an arbitrary complex constant.  From
equation~(\ref{diffalpha}), $Z$ satisfies
\begin{align} 
L_r Z &= (G+\psd J) Z \nn \\
L_\theta Z &= -(S+\psd K) Z.
\label{RZ-Zderiv}
\end{align} 
We can therefore use $Z$ to simplify a number of our equations.
Separating into modulus $X$ and argument $Y$,
\begin{equation} 
Z = X \et{ \psd Y},
\end{equation} 
we find that
\begin{align} 
L_r X &= GX \nn \\
L_\theta X &= -SX,
\end{align} 
which provides an integrating factor for $G$ and $S$.  This is
particularly important, as a comparison with the bracket
relation~(\ref{newbrckt}) now tells us that
\begin{equation} 
[XL_r, XL_\theta] = 0.
\end{equation} 
Recovering a pair of commuting derivatives like this tells us that we
should fix our displacement gauge freedom by setting $b_2=c_2=0$.
With this done, we can write
\begin{align} 
X L_r &= b(r) \dr \nn \\ 
XL_\theta &= c (\theta) \dthet,
\label{XLrXLthht}
\end{align} 
where $b(r)$ and $c(\theta)$ are arbitrary functions which we can
choose with further gauge fixing.

A search through the remaining derivatives reveals that the pair $T$
and $-H$ satisfy an equation of the form of~(\ref{intfaccon}).  We can
therefore introduce a further integrating factor, $F$, with the
properties
\begin{equation} 
L_r F = TF, \qquad L_\theta F = -H F.
\end{equation} 
Both $Z$ and $F$ are unchanged by the conjugacy operation of
equation~(\ref{conjug}).

Now that we have suitable integrating factors at our disposal, we can
considerably simplify our equations for $G+\psd J$ and $S+\psd K$ to read
\begin{align}
L_r \bigl( FZ(G+\psd J) \bigr) &= 0 \nn \\
L_\theta \bigl(FZ(S+\psd K) \bigr) &= 0 \nn \\
L_r \bigl(XZ(S+\psd K) \bigr) &= 2XZ(SG+JK) \nn \\
L_\theta \bigl(XZ(G+\psd J) \bigr) &= -2 XZ(SG+JK)
\label{newGS}
\end{align}
These equations retain the symmetry described by equation~(\ref{conjug}).
Equations~(\ref{newGS}) now focus attention on the quantity $SG+JK$.
Computing the derivatives of this, we find that
\begin{equation}
\begin{aligned}
L_r(SG+JK) &= -(G+T)(SG+JK) \\
L_\theta (SG+JK) &= (H+S)(SG+JK).
\end{aligned}
\end{equation} 
It follows that 
\begin{equation} 
L_r \bigl(XF(SG+JK)\bigr)  = L_\theta \bigl(XF(SG+JK) \bigr) = 0,
\end{equation} 
and hence that $XF(SG+JK)$ is a constant.  It turns out that the
equations can be solved if this constant is non-zero, but the
solutions require infinite disks as sources and are not asymptotically
flat in all directions.  These solutions are presented elsewhere.
Since we are interested here in asymptotically flat vacuum solutions we
must set this constant to zero, which implies that
\begin{equation}
SG+JK = 0.
\end{equation}
Equations~(\ref{newGS}) now yield
\begin{equation}
L_r \bigl( XFZ^2(G+\psd J)(S+\psd K) \bigr) = L_\theta \bigl(
XFZ^2(G+\psd J)(S+\psd K) \bigr) = 0.
\end{equation}
It follows that
\begin{equation}
XFZ^2(G+\psd J)(S+\psd K) = C_1,
\end{equation}
where $C_1$ is an arbitrary complex constant.

Remarkably, we are now in a position to separate the remaining
equations into radial and angular derivatives.  With $L_r$ and
$L_\theta$ defined by equation~(\ref{XLrXLthht}), we see that we can set 
\begin{align}
FZ(G+\psd J) &= W(\theta) \nn \\
FZ(S+\psd K) &= U(r).
\end{align}
We therefore have
\begin{equation}
XW(\theta) U(r) = C_1 F,
\end{equation}
and
\begin{align}
XZ(G+\psd J) &= C_1 / U(r) \nn \\ 
XZ(S+\psd K) &= C_1 W(\theta).
\end{align}
This provides a structure consistent with our derivatives of $G+\psd J$
and $S+\psd K$.

We next form
\begin{equation}
\frac{W(\theta)}{U(r)} = \frac{G+\psd J}{S+\psd K} = \psd
\frac{SJ-GK}{S^2+K^2}, 
\end{equation} 
which is a pure imaginary quantity.  This implies that $W$ and $U$ are
$\pi/2$ out of phase.  Since these are separately functions of $r$ and
$\theta$, their phases must be constant.  Returning to the derivatives
of $Z$ we see that
\begin{equation}	
XL_r Z = b(r) \dr Z = XZ(G+\psd J) =  C_1 / U(r)
\end{equation}
and
\begin{equation}
XL_\theta Z = c(\theta) \dthet Z = XZ(S+\psd K) = C_1 / W(\theta).
\end{equation}
It follows that $Z$ must be the sum of a function of $r$ and a
function of $\theta$.  Furthermore, these functions must also have
constant phases $\pi/2$ apart.  Since the overall phase of $Z$ is
arbitrary ($Z$ was defined up to an arbitrary complex scale factor),
we can write
\begin{equation}
Z = R(r) + \psd \Psi (\theta)
\label{newZ}
\end{equation}
where $R(r)$ and $\Psi (\theta)$ are real functions.  With $Z$ reduced
to this simple form we can solve the remaining equations by finite
power series.

\section{Power series solution}

With $Z$ given by equation~(\ref{newZ}), we now have
\begin{equation}
XZ(G+\psd J) = XL_r Z = XL_r R = R'
\end{equation}
and
\begin{equation}
XZ(S+\psd K) = -XL_\theta Z = -\psd XL_\theta \Psi = -\psd \dot{\Psi}.
\end{equation}
Here we have introduced the abbreviations 
\begin{align}	
R' &= XL_r R \nn \\
\dot{\Psi} &= XL_\theta \Psi.  
\end{align}
These help to encode the fact that $XL_r R$ is a function of $r$ only,
and $XL_\theta \Psi$ is a function of $\theta$ only.  The functions
implied in the $R'$ and $\dot{\Psi}$ notation are not necessarily pure
partial derivatives, though they can be chosen as such with suitable
gauge choices.

We now have
\begin{equation} 
\frac{F}{X} = \frac{\psd}{C_1 R' \dot{\Psi}},
\end{equation}
which implies that $C_1$ is an imaginary constant.  Since $F$ is only
defined up to an arbitrary scaling, we can choose this constant as
$i$ and write
\begin{equation}
\frac{F}{X} = \frac{1}{R' \dot{\Psi}}.
\end{equation}
Differentiating this relation now yields
\begin{equation}
X (T-G) = X \frac{X}{F} L_r \frac{F}{X} = - \frac{XL_r R'}{R'} =
- \frac{R''}{R'}
\end{equation}
and
\begin{equation}
X(H-S) = - X \frac{X}{F} L_\theta \frac{F}{X} = \frac{XL_\theta
\dot{\Psi}}{\dot{\Psi}} = \frac{\ddot{\Psi}}{\dot{\Psi}},
\end{equation}
where we have introduced the obvious notation $R''$ and
$\ddot{\Psi}$ for the second derivatives.

The final relation we need to satisfy is to construct $\alpha$ from
$R$ and $\Psi$ and equate this with $(Z_0/Z)^3$.  On writing
\begin{equation}
\alpha = (G+\psd J)^2 + (T-G)(G+\psd J) + (S+\psd K)^2 + (H-S)(S+\psd K)
\end{equation}
we see that 
\begin{align}
X^2 Z^2 \alpha 
&= {R'}^2 - R''(R+i\Psi) - \dot{\Psi}^2 - i \ddot{\Psi}(R+i\Psi) \nn \\
&= {R'}^2 -R R'' - \dot{\Psi}^2 + \Psi \ddot{\Psi} - \psd (R'' \Psi +
\ddot{\Psi} R ) .
\label{X2Z2alp}
\end{align}
This must be equated with 
\begin{equation}
X^2 Z^2 \alpha = {Z_0}^3 X^2 / Z = {Z_0}^3 Z^* = {Z_0}^3 (R-\psd \Psi)
\label{X2Z2alp2}
\end{equation}
with ${Z_0}^3$ an arbitrary complex constant.  These are the only
differential relations we need to satisfy in order to obtain a valid
solution.

We could proceed now by fixing the remaining displacement gauge
freedom and specifying forms for $R$ and $\Psi$, or $d(r)$ and
$c(\theta)$.  While this approach certainly works, we can in fact do
better by continuing to work in our abstract, intrinsic fashion.  The
key is to notice that ${R'}^2$ and $\dot{\Psi}^2$ must be power series
in $R$ and $\Psi$ respectively in order for the right-hand side of
equation~(\ref{X2Z2alp}) to be linear in $R$ and $\Psi$.  We therefore set
\begin{align}
{R'}^2 &= k_2 R^2 + k_1 R + k_0 \nn \\
\dot{\Psi}^2 &= l_2 \Psi^2 + l_1 \Psi + l_0,
\label{pwser}
\end{align}
where the $k_i$ and $l_i$ are a set of six constants (any higher order
terms vanish).  These equations are differentiated implicitly to yield
\begin{align}
R'' = k_2 R + k_1/2 \nn \\ 
\ddot{\Psi} = l_2 \Psi + l_1/2.
\end{align}
We now substitute these series into equation~(\ref{X2Z2alp}).  In order to
remove the constant and $R\Psi$ terms we must set
\begin{equation}
l_0=k_0, \qquad l_2=-k_2.
\end{equation}
The remaining terms factorise to yield
\begin{equation}
X^2 Z^2 \alpha = \half(k_1- \psd l_1)(R-\psd \Psi),
\end{equation}
which now satisfies equation~(\ref{X2Z2alp2}).  We have therefore solved
all of the equations `intrinsically'.  This process has revealed the
presence of an arbitrary complex constant controlling the Weyl
tensor, which must be understood physically.  The remaining tasks are to
pick a suitable coordinatisation and to integrate the bracket
relations to reconstruct the $g^\mu$ vectors and the metric.

\section{Integrating the bracket structure}

In arriving at an intrinsically-defined power series solution we have
completed much of the work required to generate a solution.  In
particular, we can now integrate the bracket relations~(\ref{brack}) to
solve directly for the coframe vectors, and hence for the metric.  In
so doing we have to include a number of constants of integration.
Some of these are gauge artifacts and others can be removed by
imposing either elementary or asymptotic flatness.  The end result of
this process, as we show below, is that the constant $l_1$ must be
set to zero to achieve a solution which is well-defined everywhere
outside the horizon, and asymptotically flat.  A choice of position
gauge can then be employed to set
\begin{equation}
\alp = -\frac{M}{(r-iL\cos\!\theta)^3}.
\end{equation}
This could then be used to find the coframe vectors directly, but it
is more instructive to see how the bracket structure integrates
intrinsically.

The bracket relations~(\ref{brack}) produce a series of relations for
$a_1$, $a_2$, $b_1$ and $b_2$.  By applying the known integrating
factors these become
\begin{align}
L_r \left( \frac{1}{F} a_1 \right) &= \frac{2K}{F}
\frac{a_2}{r\sin\!\theta} &
L_\theta (X a_1) &= 0 \nn \\
L_r \left(X \frac{a_2}{r\sin\!\theta}\right) &= 0 &
L_\theta \left( \frac{1}{F} \frac{a_2}{r\sin\!\theta}\right) &=
\frac{2J}{F} a_1 \nn \\
L_r \left(X \frac{d_1}{r\sin\!\theta} \right) &= 0 & 
L_\theta \left( \frac{1}{F} \frac{d_1}{r\sin\!\theta} \right) &=
\frac{2J}{F} d_2 \nn \\
L_r \left( \frac{1}{F} d_2 \right) &= \frac{2K}{F}
\frac{d_1}{r\sin\!\theta} & 
L_\theta (X d_2) &= 0.
\label{fihieqs}
\end{align}
To simplify these equations~(\ref{fihieqs}) we first form
\begin{equation}
\frac{2J}{XF} = \frac{2J}{X^2} R' \dot{\Psi} = -\frac{2\Psi
\dot{\Psi}}{X^5} {R'}^2 = L_\theta \Bigl( \frac{{R'}^2}{X^2} \Bigr) 
\end{equation}
and
\begin{equation}
\frac{2K}{XF} = \frac{2K}{X^2} R' \dot{\Psi} = -\frac{2 R R'}{X^5}
\dot{\Psi}^2 =  L_r \Bigl( \frac{\dot{\Psi}^2}{X^2} \Bigr) .
\end{equation}
The set of equations~(\ref{fihieqs}) therefore integrate simply to yield 
\begin{align}
\frac{a_1 R'}{X} - \frac{a_2 \dot{\Psi}}{X r\sin\!\theta} &= \delta_1
\nn \\
\frac{d_2 R'}{X} - \frac{d_1 \dot{\Psi}}{X r\sin\!\theta} &= \eps_1
\end{align}
where $\delta_1$ and $\eps_1$ are two arbitrary constants.  But we
know that $Xa_1$ and $Xd_2$ are independent of $\theta$, and $X
d_1/(r\sin\!\theta)$ and $Xa_2/(r\sin\!\theta)$ are independent of
$r$.  We can therefore separate the above relations and write
\begin{align}
a_1 &= \frac{\delta_0 + \delta_1 R^2}{XR'} & \frac{a_2}{r \sin\!\theta} &= 
\frac{\delta_0 - \delta_1 \Psi^2}{X \dot{\Psi}} \nn \\
d_2 &= \frac{\eps_0 + \eps_1 R^2}{XR'} & \frac{d_1}{r \sin\!\theta} &= 
\frac{\eps_0 - \eps_1 \Psi^2}{X \dot{\Psi}},
\end{align}
where $\delta_0$ and $\eps_0$ are two further constants.

The next problem to address is the behaviour of the solution on the
$z$-axis.  The problem here is the ambiguity in the coordinate system
at the two poles.  The situation can be easily sorted out by
converting back to Cartesian coordinates and demanding that the
equivalent vectors $\{g^t,g^x,g^y,g^z\}$ are well-defined on the axis.
We quickly see that on the axis we must have $a_2=0$ and $d_1=c_1$.
Similar considerations hold for the $\om_i$ bivectors.  For these we
see that $\dot{\Psi}$ must vanish on the axis and so must contain a
factor of $\sin\theta$.  It follows that
\begin{equation}
l_2 \Psi (0)^2 + l_1 \Psi (0) + l_0 = l_2 \Psi (\pi)^2 + l_1
\Psi (\pi) + l_0 = 0.
\label{b10}
\end{equation}
A further requirement from the $\om_i$ bivectors is that we must have 
\begin{equation}
\eps_0 - \eps_1 \Psi^2 - \cos\!\theta \ddot{\Psi} = 0, \qquad \theta =
0, \pi
\end{equation}
which tells us that $\ddot{\Psi}$, and hence $\Psi$, must change sign
between $0$ and $\pi$.  Since $\Psi^2$ must have the same value at $0$
and $\pi$ to ensure that $a_2$ vanishes on the axis, we see that we
must have
\begin{equation}
\Psi (0) = - \Psi (\pi).
\label{b11}
\end{equation}
We cannot have $\Psi=0$ at the poles, so equations~(\ref{b10}) and~(\ref{b11})
enforce $l_1=0$.  Solutions with a non-zero value of $l_1$ can be
constructed, but these are only valid in the upper or lower half
planes separately.  Matching these together introduces an infinite
disk of matter as a source, and the solutions are not true vacuum.
These will be discussed elsewhere.

With $l_1=0$, we now have 
\begin{equation}
\ddot{\Psi}=-k_2 \Psi.
\end{equation}
A sensible gauge choice, which produces a well-defined solution for
all $\theta$, is to set $c(\theta)=1$ in equation~(\ref{XLrXLthht}), and
scale the solution so that $k_2=1$.  It then follows that $\Psi$ is a
linear combination of $\sin\theta$ and $\cos\theta$.  Since
$\dot{\Psi}$ vanishes at the poles, we must have
\begin{equation}
\Psi (\theta) = -L \cos\!\theta, \qquad l_2=-1, \qquad l_0 = L^2.
\end{equation}
The Riemann tensor is now controlled by the complex function
$k_1(R-\psd L\cos\theta)^{-3}/2$, which determines the magnitude of the
tidal forces experienced in different directions.  A large distance
from the source we expect this to tend to the Schwarzschild value of
$-M/r^3$, so a sensible choice for the radial gauge is
\begin{equation}
R = r, \qquad k_1 = -2M.
\end{equation}
From equation~(\ref{XLrXLthht}) it follows that
\begin{equation}
b(r) = X L_r r = R',
\end{equation}
and from equation~(\ref{pwser}) we have
\begin{equation}
{R'}^2 = b(r)^2 = r^2 - 2Mr + L^2 = \Delta.
\end{equation}
Here we have introduced the standard symbol $\Delta$ for the radial
function $b(r)^2$.

With the preceding gauge choices we now have
\begin{align}
G+\psd J &= \frac{\Delta^{1/2}}{\rho(r - \psd L \cos\!\theta)}  & 
S+\psd K &= \frac{-\psd L\sin\!\theta}{\rho(r - \psd L \cos\!\theta)} \nn \\
T-G &= -\frac{r-M}{\rho\Delta^{1/2}}  &
H-S &= \frac{\cos\!\theta}{\rho\sin\!\theta},
\end{align}
where
\begin{equation}
\rho^2 = X^2 = r^2 + L^2 \cos^2\!\theta.
\end{equation}
The equation for $T$ shows that a horizon exists at $\Del=0$.  We also
now have
\begin{equation}
b_1 = \frac{\Del^{1/2}}{\rho}, \qquad c_1 = \frac{r}{\rho}.
\end{equation}
Similarly, we now have
\begin{equation}
\begin{aligned}
a_1 &= \frac{1}{\rho \Delta^{1/2}} (\delta_0 + \delta_1 r^2)   & 
a_2 &= \frac{r}{L \rho} (\delta_0 - \delta_1 L^2 \cos^2\!\theta) \\
d_1 &= \frac{r}{L \rho} (\eps_0 - \eps_1 L^2 \cos^2\!\theta) \qquad & 
d_2 &= \frac{1}{\rho \Delta^{1/2}} (\eps_0 + \eps_1 r^2) .
\end{aligned}
\end{equation}
In order that the $g^\mu$ vectors approach a flat space coordinate
frame at infinity we set $\delta_1=1$ and $\eps_1=0$.  The former
corresponds to a constant rescaling in $r$ and the latter is achieved
with a constant rotation, both of which are gauge transformations.  The
on-axis conditions $d_1=c_1$ and $a_2=0$ set $\eps_0=L$ and
$\delta_0=L^2$ respectively.  We then finally arrive at the vectors
\begin{align}
g^t &= \frac{r^2+L^2}{\rho \Delta^{1/2}} \gam^t +
\frac{L\sin\!\theta}{\rho} \gam^\phi \nn \\
g^r &= \frac{\Delta^{1/2}}{\rho} \gam^r  \nn \\
g^\theta &= \frac{1}{\rho} \gam^\theta \nn \\
g^\phi &= \frac{1}{\rho \sin\!\theta} \gam^\phi + \frac{L}{\rho
\Delta^{1/2}} \gam^t .
\end{align}
The reciprocal set is given by
\begin{align}
g_t &= \frac{\Delta^{1/2}}{\rho} \gam_t - \frac{L \sin\!\theta}{\rho}
\gam_\phi \nn \\ 
g_r &= \frac{\rho}{\Delta^{1/2}} \gam_r \nn \\
g_\theta &=  \rho \gam_\theta \nn \\
g_\phi &= \frac{(r^2+L^2)\sin\!\theta }{\rho } \gam_\phi - \frac{L
\Delta^{1/2} \sin^2\!\theta}{\rho} \gam_t.
\end{align}
The metric derived from these vectors gives the Kerr solution in
Boyer--Lindquist coordinates.

\section{Conjugate solutions}
\label{ConjSols}

At various points in the text we have pointed out the symmetry in the
equations described by equation~(\ref{conjug}).  We are now in a position
to understand how this comes about, and relate the symmetry to the
conjugation operation discussed by Chandrasekhar~\cite{cha83}.  We
first need the general version of equation~(\ref{fihieqs}) applicable when
the barred and unbarred variables are not necessarily equal.  These
are:
\begin{align}
L_r \frac{d_1}{r\sin\!\theta} &= -G \frac{d_1}{r\sin\!\theta} -(K-\Kp )
d_2  & 
L_\theta a_1 &= S a_1 - (J-\Jp ) \frac{a_2}{r\sin\!\theta} \nn \\
L_r \frac{a_2}{r\sin\!\theta} &= -G \frac{a_2}{r\sin\!\theta} - (K-\Kp )
a_1 & 
L_\theta d_2 &= S d_2 - (J-\Jp ) \frac{d_1}{r\sin\!\theta} \nn \\
L_\theta \frac{d_1}{r\sin\!\theta} &= -H \frac{d_1}{r\sin\!\theta} +
(J+\Jp ) d_2  & 
L_r a_1 &= T a_1 + (K+\Kp ) \frac{a_2}{r\sin\!\theta} \nn \\
L_\theta \frac{a_2}{r\sin\!\theta} &= -H \frac{a_2}{r\sin\!\theta} +
(J+\Jp ) a_1 & 
L_r d_2 &= T d_2 + (K+\Kp ) \frac{d_1}{r\sin\!\theta}. 
\end{align}
The full conjugacy symmetry is therefore described by
\begin{align}
a_1 &\lra \frac{d_1}{r\sin\!\theta}  & d_2 &\lra -
\frac{a_2}{r\sin\!\theta} \nn \\
L_r &\lra - L_\theta & T &\lra H \nn \\
G &\lra S & \Gp  &\lra \Sp   \nn \\
J &\lra K & \Jp  &\lra \Kp  .
\end{align}
From the form of the Riemann tensor~(\ref{Riem}) we can see that this
interchange simply swaps round the various coefficients in $\clr(B)$.
For a general, non-vacuum configuration this interchange will affect
the Einstein tensor and therefore alter the matter distribution.  For
a vacuum solution, however, the Riemann tensor reduces to the form of
equation~(\ref{genvac2}).  This is unchanged under the above
transformation, which therefore does yield a new vacuum solution.

In terms of the metric, we already know how the $g_t$ and $g_\phi$
terms transform.  The $L_r \lra - L_\theta$ interchange is achieved by
\begin{equation}
b_1 \lra -b_2/r, \qquad c_2 \lra -c_1 /r .
\end{equation}
The effect of these interchanges on the line element~(\ref{lineelm}) is to
leave the $dr^2$, $dr \, d\theta$ and $d\theta^2$ coefficients
unchanged, and to transform the remaining terms in the following
manner:
\begin{equation}
A dt^2 + 2B dt \, d\phi - C d\phi^2 \lra C dt^2 + 2B dt \, d\phi - A
d\phi^2.
\end{equation}
This is precisely the conjugation operation described by
Chandrasekhar~\cite{cha83}.  This operation produces a new vacuum
solution because the procedure can be viewed mathematically as the
complex coordinate transformation $t \lra i\phi$.  Our approach
reveals more clearly the effect of the transformation on the
various physical terms in the $\om_i$ bivectors and the Riemann
tensor.  In particular, the interchange of $G$ and $S$ is likely to be
problematic, as $S$ is required to vanish on the $z$-axis, whereas no
such restriction exists for $G$.  This is certainly true of the Kerr
solution, for which conjugacy symmetry does not yield a globally
well-defined solution.

\section{The Carter--Robinson Uniqueness Theorem}
\label{S-Unique}

One of the most remarkable features of the Kerr solution is that it
represents the unique vacuum solution to the Einstein equations which
is stationary, axisymmetric, asymptotically flat and everywhere
nonsingular outside an event horizon.  This result was finally proved
by Robinson~\cite{rob75}, building on some initial work of
Carter~\cite{car71}.  The proof proceeds in an analogous manner to
standard uniqueness proofs by constructing an integral over the space
outside the horizon.  The boundary conditions are then employed
carefully to show that a term representing the difference of two
solutions must vanish~\cite{cha83}.  The proof is highly mathematical,
so it is instructive to see how the present formalism highlights what
is going on.

So far, we have shown that the Kerr solution is the unique type~D
axisymmetric vacuum solution which is asymptotically flat and
nonsingular outside the horizon.  This proof is constructive --- we
constructed the most general solution with these properties and it
turned out to be the Kerr solution.  This is useful, as it shows that
the role of the horizon is to force the Riemann tensor to be type~D.
To see how this comes about, we must return to the general set of
equations before the restriction to a type~D vacuum was made.  These
are embodied in equations~(\ref{feqn1}), (\ref{feqn2}), (\ref{defalp}),
(\ref{remeqn}), (\ref{delbrk}) and~(\ref{xtreqn}), together with the bracket
identity~(\ref{bktidn}).

Our first task is to identify the horizon.  The definition of the
horizon is the surface over which the bivector defined by the Killing
vectors $g_t$ and $g_\phi$ is null.  The equation for the horizon is
therefore
\begin{equation}
(g_t \wdg g_\phi)^2 = 0.
\end{equation}
The volume element defined by the $\{g_\mu\}$ frame does not vanish
anywhere, so an equivalent condition is
\begin{equation}
(g^r \wdg g^\theta)^2 = 0.
\end{equation}
Since we are free to set $c_2=b_2=0$ with a choice of displacement
gauge, the horizon in the present setup is characterised by
${g^r}^2=0$, or $b_1=0$.  It follows that, at the horizon, $L_r$
acting on any finite, continuous quantity must vanish.

From the $[L_r, L_\theta]$ bracket relation~(\ref{brack}) we see that
\begin{equation}
\Gp  = - \frac{r}{c_1} L_r \Bigl( \frac{c_1}{r} \Bigr),
\end{equation}
Since $c_1$ must be finite at the horizon, it follows that $\Gp $
vanishes there.  We also have
\begin{equation}
\Sp  = \frac{1}{b_1} L_\theta b_1 = \frac{c_1}{r b_1} \partial_\theta
b_1. 
\end{equation}
But $\Sp $ cannot be singular at the horizon, as this would imply that
the Riemann tensor itself was singular.  It follows that $b_1$ must
take the form
\begin{equation}
b_1 = (r-r_0)^\eta \tilde{b}_1(r, \theta)
\end{equation}
where $r_0$ defines the horizon, which is now a surface of constant
$r$, $\eta$ is some arbitrary exponent, and $\tilde{b}_1$ is finite at
the horizon.

Next, we consider the force necessary to remain at a constant radius.
This is simplest to analyse on the $z$ axis.  An observer remaining at
rest on the $z$-axis must apply a force of magnitude $T$ in the
outward direction.  This must be singular at the horizon, otherwise it
would be possible to escape.  This argument holds at all angles, and
the horizon is therefore also defined (in this gauge) by $T\mapsto
\infty$.

The invariants in the Riemann tensor, $\alpha$ and $\delta$ must be
finite at the horizon, so $G$ and $J$ must be zero there to ensure
that the product $T(G+\psd J)$ in $\alpha$ is finite.  We also know that
$L_r \alpha$ and $L_r \delta$ must vanish at the horizon.  Since $T$
is singular there, equations~(\ref{delbrk}) force $\del=0$ at the
horizon.  It further follows from the equation for $L_r \delta$ that
$\Gp +\psd \Jp$ must vanish at the horizon (and hence are equal to their
unbarred counterparts).  Finally, since the horizon is a surface of
constant $r$, and $\delta$ vanishes there, $\delta$ must take the form
\begin{equation}
\delta = (r-r_0)^{\eta'} \tilde{\del}(r,\theta),
\end{equation}
where $\eta'$ is a positive exponent and $\tilde{\delta}$ is finite at
the horizon.  It follows that $L_\theta \del$ must also vanish at the
horizon, which in turn sets $S+\psd K=\Sp +\psd \Kp $ there.

The above argument shows that, at the horizon, the Riemann tensor is
of type~D and all the barred variables are equal to their unbarred
counterparts.  These are precisely the conditions met by the Kerr
solution.  In the light of this result, it is perhaps less surprising
that the vacuum outside a horizon is described by the Kerr solution.
Of course, this is not a complete proof of uniqueness, but it does
serve to make the result more physically natural.  One can take the
above chain of reasoning further to show that the higher derivatives
of $\delta$ also vanish at the horizon, from which it can be argued
that $\del=0$ throughout the exterior vacuum.  As shown earlier, this
alone is sufficient to restrict to the Kerr solution.

The main weakness in the preceding argument (which is shared with all
proofs of the uniqueness of the Kerr solution) is the employment of a
`bad' gauge for the solution.  The $g^\mu$ vectors and $\om_i$
bivectors are only defined outside the horizon, and various terms go
singular on the horizon.  To avoid this problem one must work with a
more general ansatz for the $g^\mu$ vectors which allows some coupling
between $t$ and $r$.  Only then can one write down globally valid
solutions.  It would certainly be useful to have a version of the
above argument which does not rely on a bad choice of gauge, and this
is a subject for future research.

\section{Conclusions}

We have introduced a new framework for the study of axisymmetric
gravitational fields.  The technique draws on results from the gauge
treatment of gravity, and builds on earlier work on
spherical~\cite{DGL98-grav} and cylindrical~\cite{DGL96-cylin}
systems.  In this paper we have concentrated on the vacuum equations.
Further publications will detail how the technique is applied to
matter configurations.  We believe that this approach can help make
progress in the long-standing problem of finding the fields inside and
outside a rotating star.  It is also expected that this framework
should be helpful in analysing the disk solutions proposed by Pichon
and Lynden-Bell~\cite{pic96}, and Neugebauer and Meinel~\cite{neu93}.
The equations with matter included will be presented elsewhere.  A
significant feature of these is that a natural set of relations for
the fluid velocity field emerge which can be carried into the vacuum
sector~\cite{chi92}.  This velocity field in turn provides a physical
basis for the Ernst equation.

The vacuum equations developed here are under-determined.  They are
only capable of unique solution when further physical restrictions are
applied.  The restriction to type~D vacua is sufficient to yield a
fully-determined set of consistent, `intrinsic' equations.  The
development of such a set is similar to the process of obtaining an
involutive set of equations in the formal theory of partial
differential equations~\cite{olv-eqv}.  The point of the latter
process is that the equations can then be solved by a power series.
In this respect, the power series solution of the equations developed
here is remarkable.  It hints at a power series solution method which
does not rely on any definite coordinatisation.  This idea warrants
serious further investigation.

We have shown that the twin restrictions of a type~D vacuum and
asymptotic flatness are sufficient to force us to the Kerr solution.
The uniqueness theorem for the Kerr solution can therefore be
interpreted as showing that the presence of a horizon forces a
restriction on the algebraic form of the Riemann tensor.  This
provides a useful physical understanding of the uniqueness of the Kerr
solution.  Some indication of how this restriction comes about can be
seen in the more general vacuum equations.  But work remains to recast
the formalism in a gauge which is not singular at the horizon.

A future goal of this approach is to streamline and refine the
techniques to the extent where time dependence can be included.  It
would then be possible to study the formation of rotating black holes,
and their spin-up under accretion.  Our approach opens up many new
possibilities for the study of such systems.

\end{document}